\documentclass[onecolumn,usenatbib]{mn2e}
\usepackage{graphicx}
\begin{document}
\title[Magnetic field decay with Hall drift]
{Magnetic field decay with Hall drift in neutron star crusts}
\author[Y. Kojima and S. Kisaka]{Yasufumi Kojima\thanks{%
E-mail:kojima@theo.phys.sci.hiroshima-u.ac.jp} 
and Shota Kisaka\thanks{%
E-mail: kisaka@theo.phys.sci.hiroshima-u.ac.jp}\\
Department of Physics, Hiroshima University, Higashi-Hiroshima, 
739-8526, Japan}
\maketitle
\begin{abstract}
The dynamics of magnetic field decay with Hall drift
is investigated. Assuming that axisymmetric magnetic 
fields are located in a spherical crust with
uniform conductivity and electron number density,
long-term evolution is calculated up to Ohmic dissipation. 
The nonlinear coupling between poloidal and toroidal components 
is explored in terms of their energies and helicity.
Nonlinear oscillation by the drift in strongly magnetized regimes
is clear only around the equipartition between two components. 
Significant energy is transferred to the poloidal
component when the toroidal component initially dominates.
However, the reverse is not true. 
Once the toroidal field is less dominant,
it quickly decouples due to a larger damping rate.
The polar field at the surface is highly distorted 
from the initial dipole during the Hall drift timescale, 
but returns to the initial dipole in a longer dissipation timescale,
since it is the least damped one.
\end{abstract}

\section{Introduction}

  The recent discovery of soft gamma repeaters (SGRs) with weak fields
\citep{2010Sci...330..944R}
has again raised problems related to magnetic field evolution
in isolated neutron stars. The activity of magnetars 
with SGRs and 
anomalous X-ray pulsars (AXPs) has been 
believed to be powered by the decay of ultrastrong magnetic fields 
\citep{1995MNRAS.275..255T,1996ApJ...473..322T}.
Their surface dipole field, which is observationally inferred from
the spin period and its time derivative, typically exceeds 
electron quantum magnetic field $ B_{Q} =4.4 \times 10^{13} $G.
The field strength of SGR0418+5729 is, however, relatively weak 
at $<7.5\times 10^{12} $G\citep{2010Sci...330..944R}.
The critical boundary between magnetars and radio pulsars 
has thus become less clear, and magnetars are not sufficiently 
characterized by their dipole field strength alone.
Their activity may be explained by hidden magnetic fields such as
poloidal components with higher-order multipoles 
or internal toroidal components. 
In either case, the field strength should be greater 
than $ B=10^{14}$-$10^{15}$G,
since other energy sources are insufficient
(see, for example, the review by \cite{2008A&ARv..15..225M}).

   The importance of Hall drift on the magnetic evolution of neutron stars
was pointed out 
before evidence of the existence of magnetars was 
available \citep{1988MNRAS.233..875J,1992ApJ...395..250G}. 
The effect, which depends on the 
field strength, becomes more important in strong regimes where 
$ B > 10^{14}$G.
Furthermore, Hall drift may induce instability under certain
conditions \citep{2002PhRvL..88j1103R}.
The effect has been considered
in analytic treatments
\citep{2004ApJ...609..999C,2007A&A...472..233R},
and as plane-parallel slab geometry
\citep{2000PhRvE..61.4422V,2003A&A...412L..33G,2004A&A...420..631R}.
These studies are useful to understand some aspects of the
mechanism, but nonlinear numerical analysis is also required to 
examine the behavior in more realistic stars. 
Several simulations have also been performed, assuming that 
the fields are located in the spherical crust region of a neutron star
\citep{1994MNRAS.266..597N,1997A&A...321..685S,
2002MNRAS.337..216H,2004MNRAS.347.1273H,2007A&A...470..303P}.
In particular, \cite{2002MNRAS.337..216H,2004MNRAS.347.1273H}
extensively studied the effect in a crust with uniform 
conductivity and density, and subsequently extended 
the discussion to stratified stellar models.
\cite{2007A&A...470..303P} also calculated the evolution 
in a realistic stratified model with a thermal history.  
These numerical calculations are based on 
spectral or quasi-spectral methods, for example expanding 
angular functions using spherical harmonics.
Limitations of such an approach are discussed therein.
For example, evolution of a purely toroidal field forms 
a steep gradient that cannot be calculated
\citep{2007A&A...470..303P}.
A recent promising approach is using a finite difference scheme to examine 
the nonlinear evolution of the Hall instability in a 2D 
slab \citep{2010A&A...513L..12P}.
Each numerical scheme has advantages and disadvantages, 
so multiple complementary approaches are needed.

  In this paper, we calculate magnetic field evolution using
a finite difference scheme to understand nonlinear 
Hall drift dynamics. The model is simplified by assuming 
an axisymmetric magnetic field located in a crust with uniform 
conductivity and electron number density.
This paper is organized as follows. In section 2, 
the model and its assumptions are described.
Magnetic field evolution is governed by an  
induction equation with nonlinear Hall drift term.
The relevant boundary conditions and 
initial configurations are also discussed. 
Sections 3 and 4 give numerical results for two distinct initial 
configurations.
One condition is a purely toroidal field in which the poloidal part is always zero 
if it is exactly zero at the initial state.
This is in contrast to a purely poloidal initial case,
for which the toroidal part is inevitably induced.
The evolution of a purely toroidal field is furthermore of interest, 
due to the similarity to Burgers' equation (e.g., \cite{1974G.B.Whitham}).
\citet{2000PhRvE..61.4422V} study the problem in a stratified 
plane-parallel slab, which corresponds to small-scale dynamics 
much smaller than the stellar radius.
The results clarify the local mechanism,
but our concern is global aspects.
How do the results change in a spherical shell like the crust? 
This problem is numerically studied
in section 3. 
Section 4 describes a second study related to the evolution of mixed 
fields with poloidal and toroidal components.
Magnetohydrodynamics simulation 
\citep{2006A&A...450.1077B,2006A&A...450.1097B,2009MNRAS.397..763B}
shows that dynamically stable configurations are such mixed ones, 
in which poloidal and toroidal field strengths are of the same order.
There is little known about the initial configuration of neutron stars 
in particular, the location, topology, and field strength. 
If both components coexist, their strengths are likely to be similar.
It is therefore important to explore energy transfer by Hall drift 
between the components over long-term evolution through Ohmic dissipation.
Section 5 presents our conclusions.

\section{Model and Formulation}
\subsection{Equations}

  The magnetic field evolution is governed by the induction 
equation with the Hall term 
\begin{equation}
\frac{\partial }{\partial t}\vec{B}= -\vec{\nabla}\times \left( \frac{c^{2}}{
4\pi \sigma }\vec{\nabla}\times \vec{B}\right) +\vec{\nabla}\times \left[ 
\frac{c}{4\pi e n_{e}}\vec{B}\times \left( \vec{\nabla}\times \vec{B}\right) 
\right] ,
 \label{Evl.eqn}
\end{equation}
where $\sigma $ is the electric conductivity, $n_{e}$ the electron number
density, $e$ the charge density, and $c$ the speed of 
light \citep{1992ApJ...395..250G,1994MNRAS.266..597N}.
There are two typical timescales associated with the first and 
second terms in eq. (\ref{Evl.eqn}), 
the Ohmic decay timescale $ 4\pi \sigma L^{2}/c^{2} $ 
and the Hall drift timescale $ 4\pi en_{e}L^{2}/(cB_{0})$. 
Here, $L$ and $B_{0}$ are typical values
for the spatial length and magnetic field strength.
In general, $\sigma $ and $n_{e}$ depend on
the spatial position and the time through the thermal history. In
this paper, however, for simplicity they are assumed to be constant to 
explore the dynamics. 
Thus, the behavior of this system is specified by 
the ratio of two timescales,
$\mathcal{R}_{m}=\sigma B_{0}/(ecn_{e} )$,
which is called the magnetized parameter 
by \cite{2007A&A...470..303P}.
The parameter $\mathcal{R}_{m}$ is given by the initial maximum 
value of the magnetic field.
The typical value for a neutron star is
$\mathcal{R}_{m}=$1-10$\times(B_{0}/10^{13}{\rm G}) $
\citep{2007A&A...470..303P},
but this significantly depends on spatial position and 
temperature \citep{2004ApJ...609..999C}. 
The overall physical timescale is scaled by
$\tau_{d}= 4\pi \sigma r_{s}^{2}/c^{2} $,
where the stellar radius $r_{s}$ is used for normalization.
Note that the characteristic decay timescale is $\tau_{d}$ for node-less 
field filled in a sphere, but it becomes smaller for the field 
localized in a crust.
A magnetic field with $\mathcal{R}_{m} (> 1)$ is initially   
given, and the evolution is numerically followed until 
decay at $\sim \tau _{d}$.
In the case where $\mathcal{R}_{m} \gg 1$, 
which is relevant to strong fields like those of magnetars, 
the second term on the right-hand side of eq. (\ref{Evl.eqn}) 
dominates. The advection term is nonlinear, and 
treating it becomes complicated.

Magnetic fields with axial symmetry ($\partial/\partial \phi=0 $) 
are described by two functions, a flux function $G$ 
describing the poloidal magnetic field and a stream 
function $S$ describing poloidal current flow:
\begin{equation}
\vec{B}=\frac{1}{R}(\vec{\nabla}G\times \vec{e}_{\phi}) 
+\frac{S}{R}\vec{e}_{\phi},
   \label{DefB.eqn}
\end{equation}%
where $R=r\sin \theta $ is the cylindrical radius in spherical coordinates
$(r, \theta, \phi)$. Ampere's equation gives the current density as 
\begin{equation}
\frac{4\pi}{c} \vec{j}= \vec{\nabla }\times \vec{B} 
 = \frac{1}{R}(\vec{\nabla}S\times \vec{e}_{\phi}) 
-\frac{1}{R}\mathcal{D}(G)\vec{e}_{\phi},
  \label{DefJ.eqn}
\end{equation}
where ${\mathcal{D}}(G)$ is given by 
\begin{equation}
{\mathcal{D}}(G) = \left( \frac{\partial ^{2}}{\partial r^{2}} 
+\frac{\sin\theta }{r^{2}}\frac{\partial } {\partial \theta } 
\frac{1}{\sin \theta }\frac{\partial}{\partial \theta } \right)G,
\end{equation}%
or in cylindrical coordinates $(R, Z, \phi)$ by
\begin{equation}
{\mathcal{D}}(G) = \left( R\frac{\partial}{\partial R} \frac{1}{R} 
\frac{\partial}{\partial R} 
+\frac{\partial ^{2}}{\partial Z^{2} }\right)G.
\end{equation}
The magnetic field evolution (\ref{Evl.eqn})
is written in terms of $G$ and $S$ (e.g., \cite{2007A&A...472..233R}) as
\begin{equation}
\frac{\partial G}{\partial t} = \frac{1}{ \tau_{d} }\mathcal{D}(G) 
+\frac{\mathcal{R}_{m}}{\tau_{d} R}
 (\vec{\nabla}G\times \vec{\nabla}S)\cdot \vec{e}_{\phi},
  \label{EvlG.eqn}
\end{equation}
\begin{equation}
\frac{\partial S}{\partial t} =\frac{1}{ \tau_{d} }\mathcal{D}(S) 
+\frac{\mathcal{R}_{m}R}{\tau_{d}}
\left[ \vec{\nabla} \times \left\{ \frac{1}{R^{2}}(\mathcal{D}(G) 
\vec{\nabla}G +S\vec{\nabla}S)\right\} \right]\cdot \vec{e}_{\phi},
\label{EvlS.eqn}
\end{equation}
where $G$, $S$, and the spatial length are appropriately normalized.
As has been pointed out (e.g., \cite{2000PhRvE..61.4422V,2007A&A...470..303P}),
the evolutionary equation of a purely toroidal
magnetic field is very similar to Burgers' equation, which is a simple
example of nonlinear propagation with diffusion in one spatial
dimension \citep{1974G.B.Whitham}.
Equation (\ref{EvlS.eqn}) for $G=0$ is thus reduced to 
\begin{equation}
\frac{\partial S}{\partial t} = \frac{1}{ \tau_{d} } 
\left( R\frac{\partial}{\partial R} \frac{1}{R} \frac{\partial}{\partial R} 
+\frac{\partial ^{2}}{\partial Z^{2} }\right)S 
+ \frac{2\mathcal{R}_{m} S}{\tau_{d} R^{2}}\frac{\partial S}{\partial Z}.
  \label{PureS.eqn}
\end{equation}
If the function $S$ depends only on $Z$, eq. (\ref{PureS.eqn}) is exactly 
Burgers' equation.

The functions $G$ and $S$ in eq. (\ref{DefB.eqn}) can be expressed by a sum
of Legendre polynomials $P_{l}(\theta)$: 
\begin{equation}
G=-\sum_{l=1}g_{l}(r,t)\sin \theta 
\frac{\partial P_{l}(\theta )}{\partial\theta },
   \label{Gexpd.eqn}
\end{equation}
\begin{equation}
S=-\sum_{l=1}s_{l}(r,t)\sin \theta 
\frac{\partial P_{l}(\theta )}{\partial\theta }.
   \label{Sexpd.eqn}
\end{equation}
The functions $g_{l}$ and $s_{l}$ independently evolve in the absence 
of the Hall term ($\mathcal{R}_{m} =0$). However, nonlinear coupling 
among $g_{l}$ and $s_{l}$ with different indices $l$ becomes important 
with the increase of $\mathcal{R}_{m} $. 
Most numerical calculations of magnetic decay with the Hall
effect have been performed by such an expansion
\citep{1994MNRAS.266..597N,
2002MNRAS.337..216H,2004MNRAS.347.1273H,2007A&A...470..303P},
with an exception\citep{1997A&A...321..685S}.
Angular part of the field is expanded by spherical harmonics, but
finite difference is used for radial direction in
quasi-spectral method\citep{1994MNRAS.266..597N,2007A&A...470..303P}. 
Tchebycheff polynomials are used for radial direction in
spectral method\citep{2002MNRAS.337..216H,2004MNRAS.347.1273H}.
The limitation of spectral or quasi-spectral methods is
also discussed, for example by \cite{2007A&A...470..303P}.
A numerical Gibbs oscillation appears when the function
evolves to form a steep gradient caused by advection. 
This is likely to occur in the large magnetized parameter $\mathcal{R}_{m} $. 
The finite difference method is used in this paper as an alternative
approach. The numerical scheme used is a simple stable one, first-order 
forward time differencing and second-order centered space (FTCS) 
differencing (see, for example, \cite{1992nrfa.book.....P}). 
There are more sophisticated schemes, but it is not easy to apply
them to the nonlinear Hall drift term, which is most important here. The
grid is staggered one with equal spacing.
Typical number of grid points is $100 \times 120$.
Our numerical results are verified using previous results based on the spectral
method in \cite{2002MNRAS.337..216H}, and the method works well for most parameters.

\subsection{Initial configuration and boundary conditions}

The magnetic field is assumed to be located outside the
superconducting core. The crust ranges from $r_{1} $ to the 
surface $r_{s}$. A typical size in neutron star models 
is $(r_{s}-r_{1})/r_{s} \sim 0.1$, and depends on the
equation of state and the stellar mass. A slightly thicker crust model is
chosen here, $r_{1}/r_{s} =0.75 $, because this allows an easier demonstration of the numerical results.
In actual implementation it is necessary to use more realistic models
that include stratified conductivity, number density, and so on,
but this simple model is useful for understanding the fundamental dynamics.

The magnetic field cannot penetrate into the core ($r < r_{1}$).
The condition for the toroidal field is simply $S=0$ at $r_{1}$. 
The condition for the poloidal field means that $G$ is a constant at $r_{1}$, 
which is chosen as $G=0$. 
As discussed in \cite{2002MNRAS.337..216H,2007A&A...470..303P},
the tangential components of the electric field, 
$E_{\theta }$ and $E_{\phi }$, should vanish. 
They are explicitly given by 
\begin{equation}
E_{(\theta , \phi)} = \frac{1}{\sigma} j_{(\theta , \phi)} +\frac{1}{e n_{e}}
( \vec{j} \times \vec{B} )_{(\theta , \phi)}.   \label{Etan.eqn}
\end{equation}
The second term on the right-hand side vanishes because 
$\vec{B} = 0$ at $r_{1}$. The first term may be negligible 
for a large conductivity $\sigma $. Otherwise, some conditions should 
be imposed to cause the tangential current to vanish. 
The condition $j_{\theta}=0$ is satisfied if
the function $S$ rapidly approaches $0$ as $r$ goes to $r_{1}$. 
The condition $j_{\phi}=0$ should be imposed on the function $G$, 
more precisely on $\mathcal{D }(G)$, which is not easily treated. 
In the numerical simulation, the conditions $G=S=0$ are used 
at $r_{1}$, assuming a large $\sigma $. 
The current distributions should be set up at the 
initial time for the conditions to be satisfied.

  The exterior of the star is assumed to be vacuum. The toroidal field should
vanish, so that the boundary condition is $S=0$ at $r_{s}$. The vacuum
solution of the poloidal field is expressed by a sum of multipole fields,
\begin{equation}
G=-\sum_{l=1} \frac{ a_{l} }{r^{l}} \sin \theta 
\frac{\partial P_{l}(\theta)}{\partial \theta }, 
  \label{Vac.eqn}
\end{equation}
where the coefficients $a_{l}$ represent the multipole moments.
For example, the dipole moment is $ a_{1}=\mu$. 
The radial component $B_{r}$ should be continuous to
the exterior vacuum solution across the surface 
by the condition ${\vec \nabla } \cdot {\vec B} =0$,
while $B_{\theta}$ may be discontinuous if a surface current is 
allowed. In this paper, neglecting the surface currents,
both components are assumed to be continuous,
and the coefficients $a_{l}$ are calculated by the 
interior numerical function $G$ at the surface $r=r_{s}$. 
The surface boundary condition can be expressed by a sum of $a_{l}$ 
up to $l_{max}=$20-30. The results slightly change if the truncation 
is $l_{max}<10$, but do not change even if $l_{max}$ is further increased.
Compared with spectral or quasi-spectral methods, small scale structure is 
not evident in our finite difference method. Initial configuration contains 
$l=$1 or 2 component, so that the truncation is justified. 
Another approach without the multipole expansion is 
proposed as non-local boundary condition by Green's 
formula\citep{2010A&A...513L..12P}, but is not used here.
Finally, for the boundary condition on the symmetric axis, $\theta =0$ 
and $\pi$ are the regularity for the functions $G$ and $S$, respectively. 
In other words, $G=0$ and $S=0$ there.

The functions $G$ and $S$ at the initial time of the numerical
simulation should also be subject to the boundary conditions.   
The configuration is easily specified by the functions $g_{l}$ 
and $s_{l}$ in eqs. (\ref{Gexpd.eqn}) and (\ref{Sexpd.eqn}). 
Initial current distribution is chosen as ${\vec j} =0$ at both inner and
outer boundaries. 
One simple solution satisfying this is given by 
\begin{equation}
s_{l}(r,0)= b_{l} \sin^{2} \left( \frac{\pi (r-r_{1})}{ r_{s}-r_{1} }
\right), 
  \label{Sinit.eqn}
\end{equation}
where $b_{l} $ is a constant. Both the function $s_{l}$ and 
its derivative $ds_{l}/dr$ approach 0
as $r \to r_{1}$ and $r \to r_{s}$.
For the poloidal field, the function $g_{l}(r,0)$
is numerically given by solving
\begin{equation}
\left( \frac{d^{2}}{dr^{2}} -\frac{l(l+1)}{r^{2}} \right)g_{l}=
 - \sin \left( \frac{\pi (r-r_{1})}{ r_{s}-r_{1} } \right),
  \label{Ginit.eqn}
\end{equation}
with boundary conditions $g_{l}(r_{1},0)=0$
and $g_{l}(r_{s},0)=a_{l}$.
The source term in eq. (\ref{Ginit.eqn}) comes from
$j_{\phi}$, which is chosen to be localized near the 
geometrical center $r \approx (r_{1}+r_{s})/2$.
The function $g_{l}(r,0)$ is smoothly connected with the multipole 
solution $a_{l}/r^{l}$ at the surface $r_{s}$.

\subsection{Energy and helicity}

In previous works (e.g., \cite{2002MNRAS.337..216H,2004MNRAS.347.1273H}), 
coefficients of Legendre polynomials
were utilized to study the system dynamics, and
are useful to understand, for example, energy transfer 
among different wavelengths at a given time.
In order to examine the whole dynamics, it is necessary to calculate 
the coefficients for many times, 
because they are time-dependent.
Our concern in this paper, however, is global aspects,
so that energy $E(t)$ and helicity $H(t)$ are used 
to represent the magnetic fields.
These are indicators of the field strength and twisted structure.
Integrating over the entire space, they  are given by
\begin{equation}
  E =\frac{1}{8\pi }\int (\vec{B}\cdot \vec{B})dV,
\end{equation}
\begin{equation}
  H =\int (\vec{A}\cdot \vec{B})dV,
\end{equation}
where $dV$ is the three-dimensional volume element,
and $\vec{A}$ is a vector potential for $\vec{B}$.
The energy is divided to poloidal $E_{P}$ and toroidal $E_{T}$
parts, $E=E_{P}+E_{T}$,  by which the energy transfer between them
is examined.  From eq. ({\ref{DefB.eqn}}),
the explicit forms are given by functions $G$ and $S$:
\begin{equation}
E_{P}=\frac{1}{8\pi }\int 
\left( \frac{ \vec{\nabla} G }{R} \right)^2 dV,
~~~
E_{T}=\frac{1}{8\pi }\int \left( \frac{ S}{ R} \right)^2 dV,
~~~
 H =2\int \frac{GS}{R^2}dV.
\end{equation}

From eqs. (\ref{Evl.eqn}) and (\ref{DefJ.eqn}), the evolution of 
magnetic energy is given by
\begin{equation}
\frac{d}{dt}E=-\int \frac{1}{\sigma } (\vec{j}\cdot \vec{j}) dV.
  \label{Edecay.eqn}
\end{equation}
This is nothing but Ohmic dissipation.
The Hall drift does not concern the energy dissipation
\citep{1992ApJ...395..250G,1994MNRAS.266..597N},
but may affect the dissipation rate
by modifying the current distribution.
Similarly, the evolution of magnetic helicity $H$ is 
\begin{equation}
\frac{d}{dt}H=-\int  2c(\vec{E}\cdot \vec{B})dV=
 -\int \frac{2c}{\sigma }(\vec{j}\cdot \vec{B})dV.
 \label{Hdecay.eqn}
\end{equation}
In this way, both energy and helicity decay through Ohmic 
dissipation. Note that $E$ and its dissipation term, 
the right-hand side of eq. (\ref{Edecay.eqn}), have a definite sign, 
but $H$ and its dissipation term, the right-hand side of  
eq. (\ref{Hdecay.eqn}) do not.
The energy should monotonically decrease, but evolution of the helicity 
is unknown without numerical integration. 
In the numerical calculations, the energy balance equation (\ref{Edecay.eqn}) 
is used to check the accuracy.  The relative errors to the initial 
magnetic energy are less than $10^{-2}$.
They decrease with increase of the grid number. 
Relatively large numerical error is produced in the simulation with 
large magnetized parameter $\mathcal{R}_{m}$. 
It is also prominent in the initial evolution, in which steep 
structure is formed. Functions in the late phase are rather smooth, 
so that the errors do not accumulate so much, even if numerical 
integration continues to a longer time.

\section{Evolution of purely toroidal fields}

This section discusses the evolution of purely toroidal magnetic fields.
The system is similar to that of Burgers' equation, as discussed in Section 2. 
The initial configuration of $S$ is specified by a single 
component $s_{l}(r,0)$ in eq. (\ref{Sinit.eqn}). 
The three models considered here are characterized by 
$l=1$ with $b_{1} >0$ in Model A, 
$l=2$ with $b_{2} >0$ in Model B, and 
$l=2$ with $b_{2} <0$ in Model C. 
The toroidal magnetic field $B_{\phi}$ in Model C has the same 
amplitude as Model B, but opposite direction. 
These initial configurations are shown in 
the top to bottom panels of Fig. 1.
The coefficient $b_{1}$ or $b_{2}$ is chosen for the maximum 
of $|B_{\phi}|$ to be equal to 1, since the
magnetic fields in the numerical calculation are normalized by the maximum
value at the initial state. 
The magnetized parameter for these models is $\mathcal{R}_{m}=100$.

  The top row shows the evolution of $l=1$ with $b_{1} >0$ (Model A). 
Snapshots of the configuration are shown at the initial state (left),
at the Hall drift timescale $ \sim 6 \times 10^{-4}\tau_{d}$ (center)
and at the Ohmic decay timescale $ \sim 2 \times 10^{-3}\tau_{d}$ (right).
Note that the amplitude of $s_{1}$ decays 
by $\exp(-166t/\tau_{d})$ in the absence of the Hall 
term \citep{2002MNRAS.337..216H}, so the
typical decay time is not $\tau_{d}$ but 
$ 6\times 10^{-3} \tau_{d}$. 
The difference mainly comes from the choice of normalization length.
The stellar radius is used in this paper to compare previous works, but 
more appropriate one is the crust size.
Thus, $\tau_{d}$ is rather large to characterize the actual Ohmic decay.
The maximum of $S$ at $t=0$ is located at $r\approx (r_{1}+r_{s})/2$ and 
$\theta=\pi/2$, and moves in the negative $Z$-direction, 
as shown in close-ups in Fig.1.
The drift stops at the outer boundary in several times the drift timescale.
It is clear that the function $S$ at this time
contains higher multipole components $s_{l}(r,t)$ in addition to 
the initial $s_{1}(r,t)$
when it is expanded as in eq. (\ref{Sexpd.eqn}). 
Subsequently the shape is nearly fixed, but overall amplitude 
gradually decreases with a longer Ohmic decay timescale. 
Some numerical calculations were performed longer until $\sim \tau_{d}$, 
but the behavior after $ 6 \times 10^{ -2} \tau_{d}$ is well 
described by a simple exponential decay.
The results are limited to the early phase since subsequent 
evolution is easily inferred from the extrapolation.

  The middle row shows the evolution of Model B ($b_{2} >0$), 
and the bottom row shows the evolution of Model C ($b_{2} <0$).
The evolution of the magnetic configuration is quite different
according to the initial sign of $b_{2}$. 
The positive and negative regions of $S$
`collide' at the equatorial plane in Model B, 
while they `repulse' each other in Model C. 
This occurs because the drift is negative in the $Z$-direction for $S>0$, 
but positive in the $Z$-direction for $S<0$. 
It is also important to note that the shape predominantly moves 
in not the $\theta$-direction, but the $Z$-direction 
(see eq. (\ref{PureS.eqn})).
The motion in $S$ is clear in close-ups of steep structure in Fig.1.
In any model, the moved shape decays in a longer decay timescale.

These results remarkably show the nature of the Hall drift, where  
a different initial sign for $B_{\phi}$ leads to a different fate. 
This leads to an interesting question: 
Do these different configurations lead to significantly different 
dissipation rates?
If so, the direction of $B_{\phi}$ would have a significant effect. 
Figure 2 shows magnetic energy time evolutions for three models. The
difference between Model B and Model C is relatively minor.
A more marked difference is shown for 
the initial multipole. That is, the energies for 
Model B and C decay slightly faster than that of Model A. 
The current is swept to the outer boundary in Model C, but 
the outer boundary is replaced by the equatorial plane in Model B. 
In Fig. 2, a curve with $\exp(-332 t/\tau_{d})$, 
corresponding to the energy decay in the absence of the Hall drift, 
is also plotted for comparison.

  Figure 3 shows the dependence of the magnetic parameter ${\mathcal R}_{m}$. 
The initial configuration is the same as that of Model B, 
the model of $l=2$ with $b_{2}>0$. The same figure also shows the magnetic 
energy for free decay. The damping is clearly strong for 
large ${\mathcal R}_{m}$. The Ohmic decay is enhanced in the presence of
the Hall drift, that is, current is swept into a certain region, where 
it is effectively dissipated.

\begin{figure}
  \includegraphics[scale=1.0]{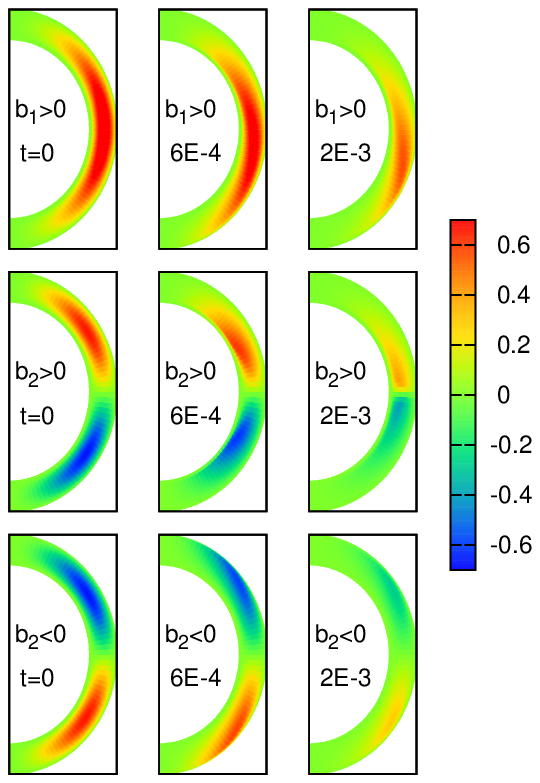} %
\hspace{-20mm}
%
  \includegraphics[scale=1.0]{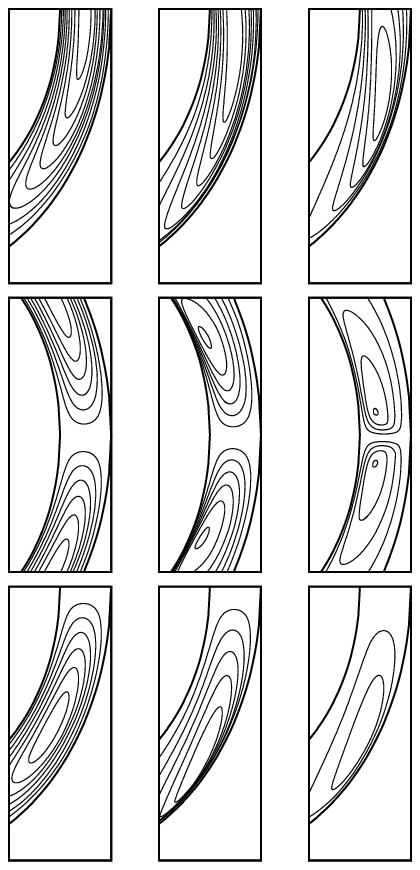} %
\caption{ Snapshots of time evolution for the function $S$, 
Model A ($b_{1} >0$) at top, Model B ($b_{2} >0$) in the middle,
and Model C ($b_{2} <0$) at bottom.
Column times are $t/\tau_{d} =0, 6 \times 10^{-4}, 2 \times 10^{-3}$ 
from left to right. 
Whole spherical regions are shown by color contour 
in the left nine panels, while their close-ups of the steep region 
are shown by contour line with increment of 0.1 in the right nine panels. 
The region is limited to 
$0.5 \le R/r_{s} \le 1, -1 \le Z/r_{s} \le 0$ in Model A,
$0.5 \le R/r_{s} \le 1, -0.5 \le Z/r_{s} \le 0.5$ in Model B,
and  $0.5 \le R/r_{s} \le 1, -1 \le Z/r_{s} \le 0$ in Model C.
}
\end{figure}
%

\begin{figure}
\centering
\includegraphics[scale=0.75]{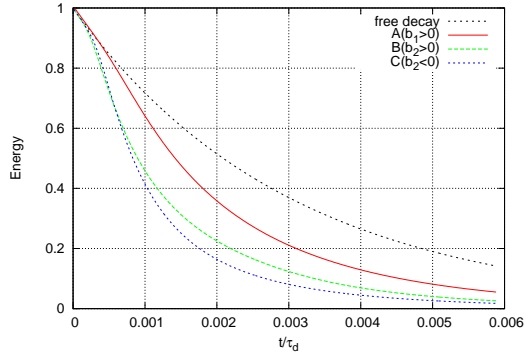}
\caption{ Normalized magnetic energy as a function of 
time $t/\protect\tau_{d}$ for three models with magnetization 
parameter $\mathcal{R}_{m} =100$.
Free decay curve is also plotted for comparison. }
\end{figure}
\begin{figure}
\centering
\includegraphics[scale=0.75]{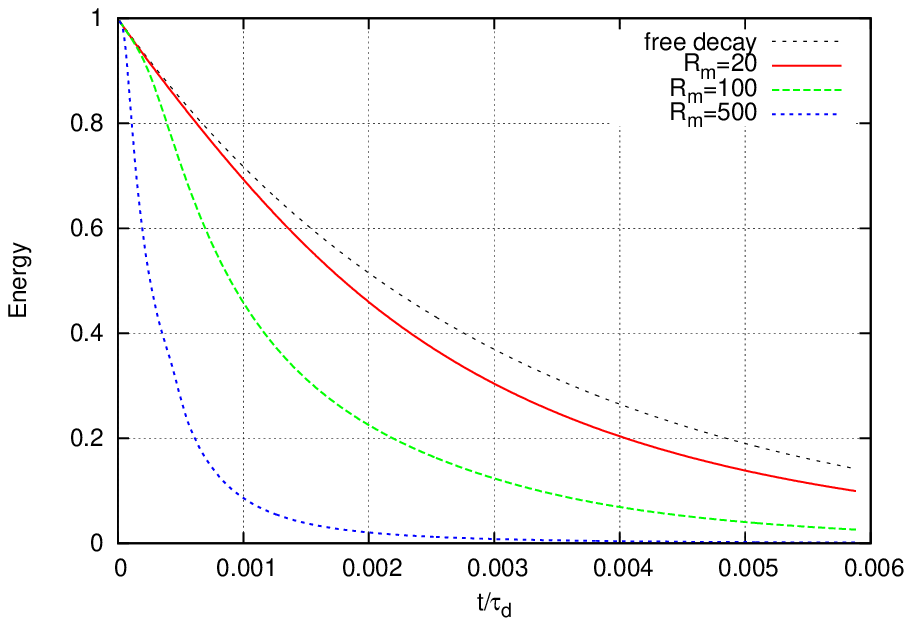}
\caption{ Normalized magnetic energy as a function of time 
$t/\protect\tau_{d}$ for $\mathcal{R}_{m} =20,100, 500$. 
Free decay curve is also plotted for comparison. }
\end{figure}

\section{Evolution of mixed fields}

In this section, field evolution is numerically studied for a mixed 
magnetic configuration consisting of poloidal and toroidal fields.
The initial configuration is given solely by the $l=1$ component for
both fields, namely, eq. (\ref{Sinit.eqn}) for $s_{1}$
and eq. (\ref{Ginit.eqn}) for $g_{1}$. 
The maximum of each field is chosen as the same amplitude, 
and the magnetized parameter is $\mathcal{R}_{m}=100$.
Figure 4 shows snapshots of the evolving fields
at representative times. The color contour represents 
the function $S$ of the toroidal field, and lines denote
the contour of the magnetic flux function $G$ of the 
poloidal field.

  Oscillatory behavior is clearly evident in $G$.
Initially, the function decreases
with the increase in cylindrical distance, 
and the maximum is located on the equator $\theta=\pi/2$.
The maximum moves `upward' in the meridian plane,
toward $\theta<\pi/2$, until 
$t/\tau_{d} \approx 1.4 \times 10^{-3}$ (second panel). 
It then changes direction and goes `downward,' passing through 
the equator at $t/\tau_{d} \approx 3.2 \times 10^{-3}$ (third panel) 
and reaching a minimum 
at $t/\tau_{d} \approx 5.2 \times 10^{-3}$ (fourth panel), 
before returning to the initial position at 
$t/\tau_{d} \approx 7.8 \times 10^{-3}$ (fifth panel).   
During this cycle, the field strength decreases.

  The function $S$ is also oscillatory.
The initial configuration contains only the $l=1$ component
in the angular part ($ S \propto \sin^2 \theta $), which is
symmetric with respect to $\theta=\pi/2$.
The state at
$t/\tau_{d} \approx 1.4 \times 10^{-3}$ (second panel) 
markedly differs from the initial state.
The configuration is no longer symmetric, and higher multipoles 
can be seen. The field strength itself is weak around this time.
At $t/\tau_{d} \approx 3.2 \times 10^{-3}$ (third panel), 
the configuration again becomes symmetric like the initial state,
but the sign of $S$ is reversed. The $l=1$ component is dominated 
there. After the direction of $B_{\phi}(=S/(r\sin\theta))$ again
changes, the configuration returns to the initial one at
$t/\tau_{d} \approx 7.8 \times 10^{-3}$ (fifth panel).   
The directional change occurs around 
$t/\tau_{d} \approx 1.4 \times 10^{-3}$ (second panel)
and $5.2 \times 10^{-3}$ (fourth panel), 
which correspond to a local minimum of toroidal field strength.
The overall toroidal field strength also decreases during 
this cycle.

\begin{figure}
\centering
\includegraphics[scale=1.0]{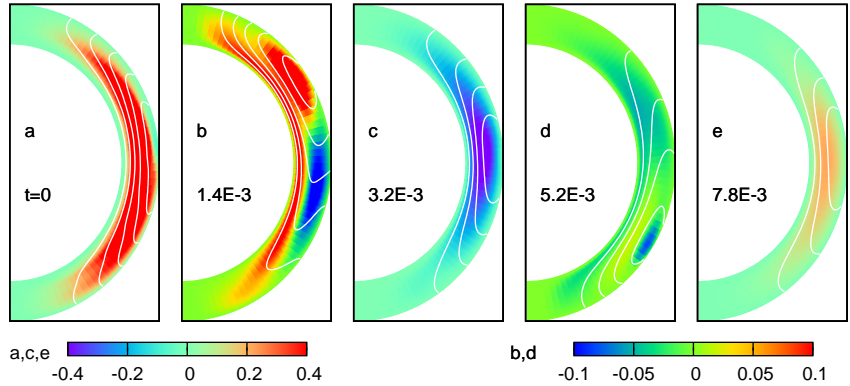} %
\caption{Snapshots of time evolution for functions $G$ and $S$. 
Contour lines outwardly represent the level of $G$
for $0.02 \times n \times (B_{0} r_{s}^{3}), n=1,2,\cdots$.
Color contour represents $S$ normalized by
$B_{0} r_{s}$.
Note that different color scales are used,
since $S$ becomes very small at the bounces in the second
and fourth panels. 
Those panels use the color scale on the left; the others, the scale on the right.
}
\end{figure}

Figure 5 clearly shows the oscillatory behavior of the
magnetic energy, which is divided into poloidal and toroidal 
parts, $E_{P}$ and $E_{T}$, respectively. 
The magnitude of the amplitudes at $t=0$ are nearly the same.
\footnote{
Values are not exactly the same, because the maximum amplitude of 
each field is fixed at same value, the distributions are slightly different.}
The curve of toroidal energy $E_{T}$ represents a damped oscillation.
Local minima can be seen at $t/\tau_{d} \approx 1.4 \times 10^{-3}$
and $5.2 \times 10^{-3}$,
corresponding to the time of the second and fourth panels in Fig. 4.
The configurations of the third and fifth panels in Fig. 4 are those of
the local maxima at $t/\tau_{d} \approx 3.2 \times 10^{-3}$
and $7.8 \times 10^{-3}$.
There is remarkable energy transfer between the toroidal and poloidal parts
during the initial phase. Initially $E_{P}$ increases 
until $t/\tau_{d} \approx 1.4 \times 10^{-3}$, although 
the total energy $E_{sum}=E_{P}+E_{T}$ decreases.
The sum always decreases due to Ohmic decay
(see eq. (\ref{Edecay.eqn})). 
The initial rapid decay of $E_{T}$ is thus partially due to this 
transfer. Energy is subsequently transferred between the two components 
in turn, but the behavior becomes less clear.
The magnetic field decays on the timescale 
$t/\tau_{d} \approx 10^{-2}$, so the coupling becomes weak.
The oscillation period gradually becomes longer,  
since the drift timescale increases.
Figure 5 also shows the evolution of magnetic helicity $H$,
which exhibits oscillatory damping with Ohmic decay timescale
$t/\tau_{d} \approx 10^{-2}$.
The change of sign in $H$ denotes an inversion of the toroidal 
field $B_{\phi }$, which occurs around the local minima of $E_{T}$.

The coefficients $a_{l}$ 
of the multi-moments in eq. (\ref{Vac.eqn}) describe the exterior poloidal field.
The right panel of Fig. 5 shows the evolution of a few of the lowest values.
Higher multipoles are induced until 
$t/\tau_{d} \approx 1.4 \times 10^{-3}$, 
the configuration shown in the second panel of Fig. 4.
The poloidal field is no longer dipole at this time.
Interestingly, the coefficient $a_{1}$ is significantly decreased, 
as compared with the initial value, 
although a large amount of energy is stored in the poloidal part, 
as shown in the left panel of Fig. 5.
Dipole field strength at this time is not a good indicator of
overall magnetic energy. 
The poloidal field returns back to the dipole at 
$t/\tau_{d} \approx 3.2 \times 10^{-3}$
(the time of the third panel in Fig. 4), 
at which higher multipoles are temporarily zero.
After sinusoidal oscillation, higher multipoles decay rather rapidly, 
so the polar dipole remains for the later time
$t/\tau_{d} > 10^{-2}$.
The behavior after $10^{ -2}\tau_{d}$ is rather simple.
Magnetic energy is significantly dissipated, so that the Hall drift 
becomes less important. Subsequent evolution in the late phase is described 
by a free decay due to Ohmic dissipation, and is omitted here.

\begin{figure}
\centering
\includegraphics[scale=1.0]{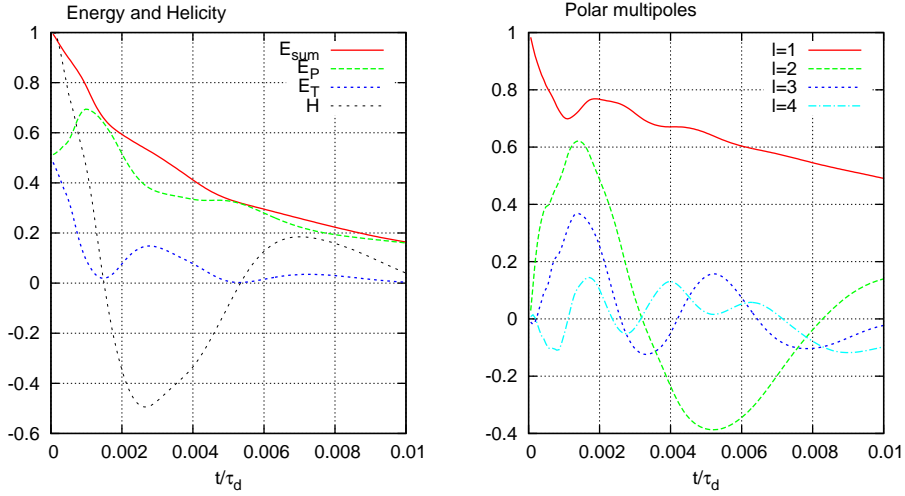}
\caption{ Time evolution of energy $E$, helicity $H$ (left panel), 
and multipole moments $a_{l}$ at the surface (right panel).
Energy is normalized by the initial total energy, and helicity
by the initial value. Coefficient $a_{l}$ of the $l$-th moment 
is normalized by the initial dipole value $a_{1}r_{s}^{l-1} $.}
\end{figure}

  Figures 6--9 compare evolutions of energy, helicity, and polar multipole 
coefficients for four models.
The initial magnetic configuration is the same as that of Figs. 4 and 5, 
but the strength and magnetized parameter $\mathcal{R}_{m}$ 
vary. The ratio of poloidal to toroidal fields is $1/4,$ or $4$ in amplitude, 
so $16(=4^{2})$ or $1/16$ in energy,
and $\mathcal{R}_{m}$ is $\mathcal{R}_{m}=20$, or $100$.
The right panel shows energy plotted on a logarithmic scale, $\log_{10}(E)$,
and the left helicity $H$ and a few lowest multipole coefficients $a_{l}$.
Figure 6 shows the results of a model with $\mathcal{R}_{m}=20$, 
in which the energy is initially dominated by the poloidal part. 
Polar multipole components are induced through the coupling to
the toroidal part, but the dipole field is barely affected.
The decay curve is well described by $\exp(-55 t/\tau_{d})$,
which coincides with free decay of the dipole.
The energy is always dominated by the polar dipole,
and the evolution is described by $\exp(-110 t/\tau_{d})$,
a line in the logarithmic scale. The curve of toroidal energy is 
nonlinear on the logarithmic scale, but is oscillatory.
The minor component is highly affected by the poloidal one, but
is almost neglected in the dynamics of the whole system.
No consequence of the toroidal field comes from the
fast decay rate $\exp(-166 t/\tau_{d})$ in the amplitude
as seen in Section 4, or from the initial small strength.

  Figure 7 shows the results of an evolution in which the toroidal 
component is dominated at the initial state. Compared with the decay 
curve of total energy in Fig. 6, the damping is much faster. 
Most of the magnetic energy is initially stored in the toroidal part, 
but rapidly decays. The poloidal component decays 
rather slowly, and dominates at the later time. The higher 
multipole moments are induced 
and have larger amplitudes than those of Fig. 6.
The overall dipole decay curve is nonetheless very similar to that 
of the large poloidal case.

  Figure 8 shows the results for large a magnetization 
parameter $\mathcal{R}_{m}=100$ and an initially large poloidal field.  
The evolution of the total energy is very similar to that of Fig. 6,
although the oscillatory behavior in the toroidal energy is evident. 
The timescale determined by the Hall drift is approximately 1/5 
that of Fig. 6, in which $\mathcal{R}_{m}=20$.
The oscillatory behavior is also clear in the coefficients $a_{l}$
and helicity $H$. However, except for the initial wavy 
structure seen in the Hall drift timescale,
the decay of the dipole field is similar to that of Fig. 6. 
The dominant component, the polar dipole, is thus not affected by
the toroidal field, 
which never plays an important role because of its rapid decay.

  Figure 9 shows the results for a large magnetization 
parameter $\mathcal{R}_{m}=100$ and an initially large toroidal field.  
In this case, a large amount of energy is transferred from the
toroidal to poloidal components.
Higher multipoles in the poloidal field are induced
around $ t/\tau_{d}\approx 10^{-3}$. 
The configuration of the poloidal field at this time is significantly 
distorted as in the second panel in Fig. 4.
The toroidal energy becomes much smaller than the poloidal energy
around $ t/\tau_{d}\approx 10^{-2}$, and 
both components are subsequently decoupled. After that, the dipole 
field evolution is determined by free decay. 
Despite the disorder at $ t/\tau_{d}\approx 10^{-3}$,
the amplitude of the free decay phase, for example 
at $ t/\tau_{d} = 2 \times 10^{-2}$,
does not differ so much from that of Figs. 6--8.

\begin{figure}
\centering
 \includegraphics[scale=0.75]{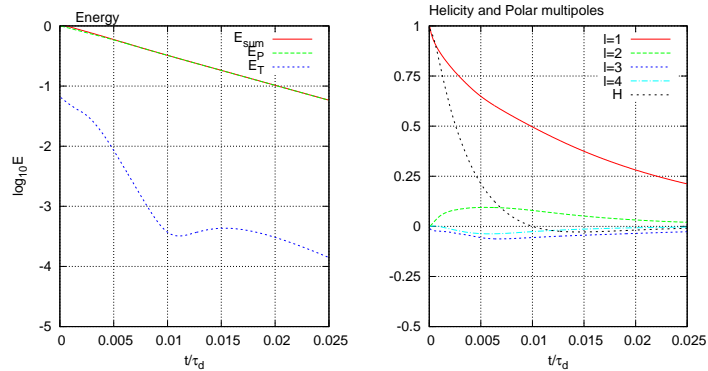}
\caption{ Time evolution of energy in logarithmic scale $\log_{10}(E)$
(left panel) and helicity $H$ and multipole moments $a_{l}$ 
at the surface (right panel) for an initially large poloidal field
with ${\mathcal R}_m=20$. 
These normalizations are the same as in Fig. 5.}
\end{figure}
\begin{figure}
\centering
 \includegraphics[scale=0.75]{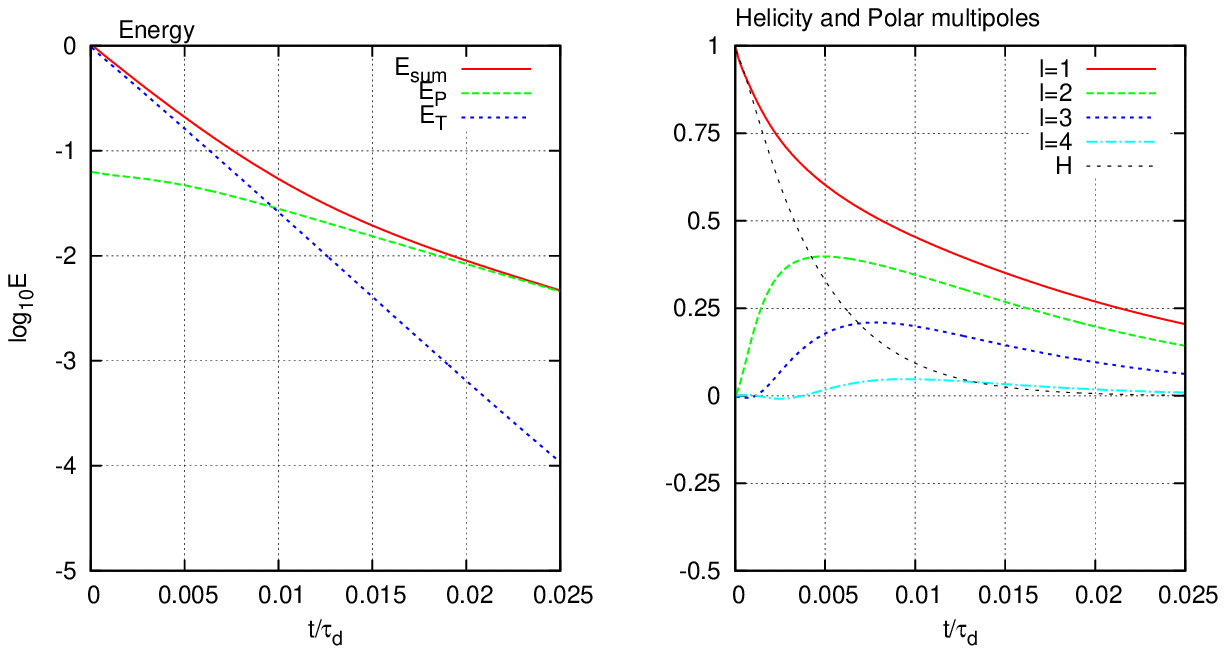}
\caption{ Same as Fig. 6, but for an initially large toroidal field
with ${\mathcal R}_m=20$. }
\end{figure}
\begin{figure}
\centering
\includegraphics[scale=0.75]{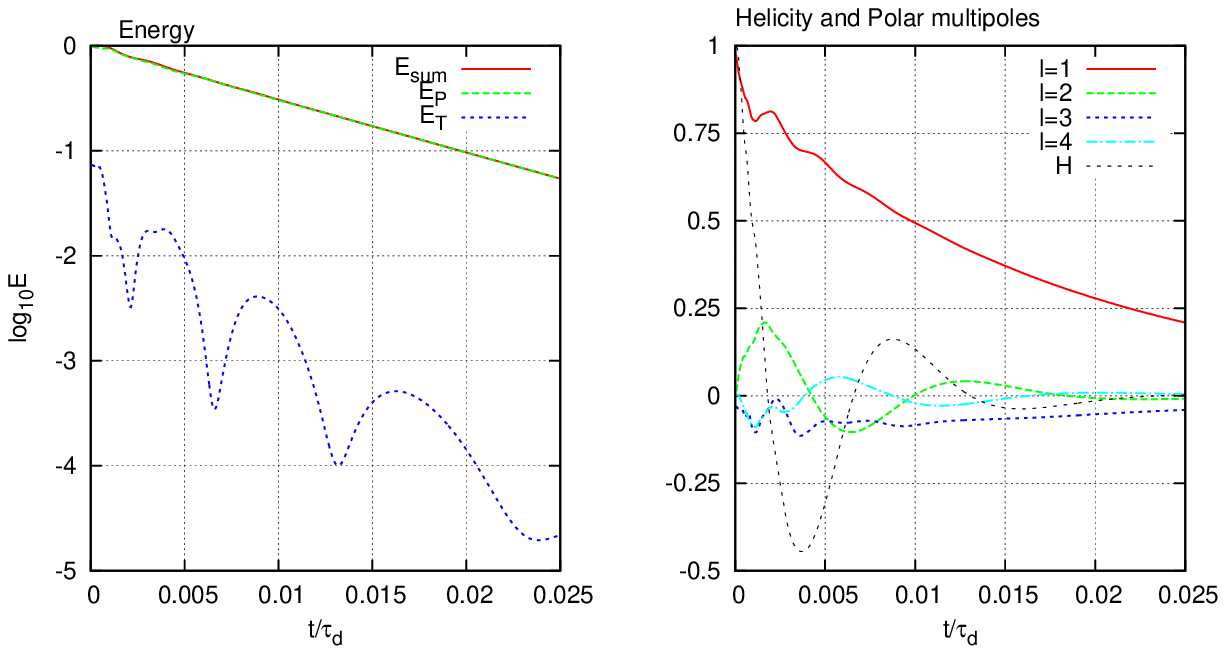}
\caption{ Same as Fig. 6, but for an initially large poloidal field
with ${\mathcal R}_m=100$. }
\end{figure}
\begin{figure}
\centering
\includegraphics[scale=0.75]{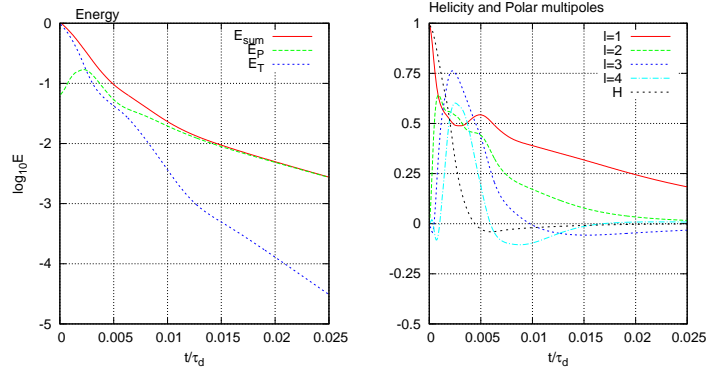} 
\caption{ Same as Fig. 6, but for an initially large toroidal field
with ${\mathcal R}_m=100$. }
\end{figure}

\section{Summary and Discussion}

  Numerical simulations demonstrate how the Hall drift 
changes the current and magnetic configuration.
In a purely toroidal evolution, Ohmic dissipation is enhanced
by accumulated currents elsewhere. 
The spatial location depends on the initial data,
but the energy dissipation rate does not so significantly depend 
on the accumulation position in our uniform conductivity model.
In a realistic case the conductivity will decrease with the radius,
so most of the magnetic energy may be effectively dissipated 
near the surface. The results become more sensitive to the initial 
condition. 
Energy is, in principle, transferred between poloidal and toroidal 
components if both are initially involved.
It is important for understanding the evolution
that the free decay rate of the polar dipole is the least.
When the polar dipole dominates, it is rarely affected by
the toroidal field, which rapidly decays.
On the other hand, a significant amount of energy is transferred 
to the poloidal field until almost equipartition
when the toroidal component is initially dominant. 
Global nonlinear coupling is manifest in the Hall drift timescale 
only when the corresponding energies are of the same order.
Moreover, the poloidal field at the surface and the exterior 
is highly distorted, and is no longer described by a pure dipole 
field. During this phase, the dipole is not
a good indicator of overall field strength.  
In a longer Ohmic dissipation timescale the toroidal field 
decays rapidly, so the coupling vanishes.
The polar dipole eventually survives.

  The early evolution highly depends on the choice of the
initial data, i.e, the configuration and strength.
As discussed by \cite{2007A&A...472..233R},
stationary conditions in the presence of Hall drift 
for an axisymmetric magnetic field means that
the isosurface of $S$ coincides with that of $G$; in other words, $S=S(G)$,
for which the coupling term ${\vec \nabla G}\times {\vec \nabla S}$
vanishes. 
This condition is not satisfied in the whole shell region.
The outer boundary condition requires that the toroidal component is
concentrated in the interior ($S=0$), whereas
the poloidal one may leak out to the exterior, meaning 
that $G \ne 0$ at the surface.
The interior poloidal field is described by the value at the surface 
if the topology is simply connected. 
The condition $S(G)$ therefore means that $S=0$ everywhere,
irrespective of $G$.
Our numerical calculation starts with an initial 
configuration different from the `Hall equilibrium', 
and shows the behavior to the state $S=0$.
It may be necessary as a next step to study plausible initial
configurations, since little known about them.

  The timescale is less accurate in our simplified model
with uniform density and conductivity.
Realistic models are necessary, but the distributions depend on 
equation of state and neutron star mass. 
Assuming that our uniform model is obtained as a result of 
spatial average of a certain model with stratified number density 
and conductivity, the timescale is estimated.
The overall normalization constant $\tau _{d} $ is given by
$\tau _{d}= 4\pi \sigma r_{s}^{2}/c^{2}$
$\approx 4.4\times 10^{9}$
$( \bar{\sigma} /10^{25} \mathrm{s}^{-1})$
$(r_{s}/10 \mathrm{km}) ^{2}$ years, where
$\bar{\sigma} $ is an averaged value of the conductivity.
The decay timescale of the dipole is
$2\times10^{-2}\tau _{d}\approx 9\times 10^{7}$ years,
and that of the toroidal one is
$6\times10^{-3}\tau _{d}\approx 3\times 10^{7}$ years.
These numbers are slightly larger, since the crust size 
in our present model is thick, $L= r_{s}/4$. 
The timescale is proportional to $L^2$, so that 
the actual values may be smaller by a factor of
$\sim 10^{-2}$--$10^{-1}$.
A typical decay timescale is around $10^{5}$--$10^{7}$ years.

  The characteristic age of the low-field magnetar SGR0418+5729 
is more than $ 2.4\times 10^{7} $ years \citep{2010Sci...330..944R}.
The dipole field ($<7.5\times 10^{12} $G) may decay or survive
within this period. However, the internal toroidal field is 
likely to dissipate more quickly.
It is therefore difficult to understand why the toroidal field has 
a much larger field strength, $>10^{13}$-$10^{14}$G, 
which is required for the activity.
One possible explanation is that current age estimates are inaccurate.
The characteristic age is normally estimated by
assuming a constant dipole magnetic field. 
The Hall drift significantly affects the surface value,
especially at the initial epoch with strong field.
As demonstrated in the second panel of Fig. 4, the surface field is 
highly distorted from the pure dipole.
SGR0418+5729 may correspond to a young phase of 
oscillatory evolution 
in which the surface dipole temporarily decreases, but
there is a strong internal toroidal component.

\section*{Acknowledgements}
This work was supported in part by a Grant-in-Aid for Scientific Research
(No.21540271) from the Japanese Ministry of Education, Culture, Sports,
Science and Technology(Y.K.), and from the Japan Society for Promotion 
of Science (S.K.). 

\begin{thebibliography}{}

\bibitem[\protect\citeauthoryear{{Braithwaite}}{{Braithwaite}}{2009}]{2009MNRA%
S.397..763B}
{Braithwaite} J.,  2009, MNRAS, 397, 763

\bibitem[\protect\citeauthoryear{{Braithwaite} \& {Nordlund}}{{Braithwaite} \&
  {Nordlund}}{2006}]{2006A&A...450.1077B}
{Braithwaite} J.,  {Nordlund} {\AA}.,  2006, A\&A, 450, 1077

\bibitem[\protect\citeauthoryear{{Braithwaite} \& {Spruit}}{{Braithwaite} \&
  {Spruit}}{2006}]{2006A&A...450.1097B}
{Braithwaite} J.,  {Spruit} H.~C.,  2006, A\&A, 450, 1097

\bibitem[\protect\citeauthoryear{{Cumming}, {Arras} \& {Zweibel}}{{Cumming}
  et~al.}{2004}]{2004ApJ...609..999C}
{Cumming} A.,  {Arras} P.,    {Zweibel} E.,  2004, ApJ, 609, 999

\bibitem[\protect\citeauthoryear{{Geppert}, {Rheinhardt} \& {Gil}}{{Geppert}
  et~al.}{2003}]{2003A&A...412L..33G}
{Geppert} U.,  {Rheinhardt} M.,    {Gil} J.,  2003, A\&A, 412, L33

\bibitem[\protect\citeauthoryear{{Goldreich} \& {Reisenegger}}{{Goldreich} \&
  {Reisenegger}}{1992}]{1992ApJ...395..250G}
{Goldreich} P.,  {Reisenegger} A.,  1992, ApJ, 395, 250

\bibitem[\protect\citeauthoryear{{Hollerbach} \& {R{\"u}diger}}{{Hollerbach} \&
  {R{\"u}diger}}{2002}]{2002MNRAS.337..216H}
{Hollerbach} R.,  {R{\"u}diger} G.,  2002, MNRAS, 337, 216

\bibitem[\protect\citeauthoryear{{Hollerbach} \& {R{\"u}diger}}{{Hollerbach} \&
  {R{\"u}diger}}{2004}]{2004MNRAS.347.1273H}
{Hollerbach} R.,  {R{\"u}diger} G.,  2004, MNRAS, 347, 1273

\bibitem[\protect\citeauthoryear{{Jones}}{{Jones}}{1988}]{1988MNRAS.233..875J}
{Jones} P.~B.,  1988, MNRAS, 233, 875

\bibitem[\protect\citeauthoryear{{Mereghetti}}{{Mereghetti}}{2008}]{2008A&ARv.%
.15..225M}
{Mereghetti} S.,  2008, A\&AR, 15, 225

\bibitem[\protect\citeauthoryear{{Naito} \& {Kojima}}{{Naito} \&
  {Kojima}}{1994}]{1994MNRAS.266..597N}
{Naito} T.,  {Kojima} Y.,  1994, MNRAS, 266, 597

\bibitem[\protect\citeauthoryear{{Pons} \& {Geppert}}{{Pons} \&
  {Geppert}}{2007}]{2007A&A...470..303P}
{Pons} J.~A.,  {Geppert} U.,  2007, A\&A, 470, 303

\bibitem[\protect\citeauthoryear{{Pons} \& {Geppert}}{{Pons} \&
  {Geppert}}{2010}]{2010A&A...513L..12P}
{Pons} J.~A.,  {Geppert} U.,  2010, A\&A, 513, L12

\bibitem[\protect\citeauthoryear{{Press}, {Teukolsky}, {Vetterling} \&
  {Flannery}}{{Press} et~al.}{1992}]{1992nrfa.book.....P}
{Press} W.~H.,  {Teukolsky} S.~A.,  {Vetterling} W.~T.,    {Flannery} B.~P.,
  1992, {Numerical recipes in FORTRAN. The art of scientific computing}.
{Cambride University Press, New York}

\bibitem[\protect\citeauthoryear{{Rea}, {Esposito}, {Turolla}, {Israel},
  {Zane}, {Stella}, {Mereghetti}, {Tiengo}, {G{\"o}tz}, {G{\"o}{\u g}{\"u}{\c
  s}} \& {Kouveliotou}}{{Rea} et~al.}{2010}]{2010Sci...330..944R}
{Rea} N.,  {Esposito} P.,  {Turolla} R.,  {Israel} G.~L.,  {Zane} S.,  {Stella}
  L.,  {Mereghetti} S.,  {Tiengo} A.,  {G{\"o}tz} D.,  {G{\"o}{\u g}{\"u}{\c
  s}} E.,    {Kouveliotou} C.,  2010, Sci., 330, 944

\bibitem[\protect\citeauthoryear{{Reisenegger}, {Benguria}, {Prieto}, {Araya}
  \& {Lai}}{{Reisenegger} et~al.}{2007}]{2007A&A...472..233R}
{Reisenegger} A.,  {Benguria} R.,  {Prieto} J.~P.,  {Araya} P.~A.,    {Lai} D.,
   2007, A\&A, 472, 233

\bibitem[\protect\citeauthoryear{{Rheinhardt} \& {Geppert}}{{Rheinhardt} \&
  {Geppert}}{2002}]{2002PhRvL..88j1103R}
{Rheinhardt} M.,  {Geppert} U.,  2002, Phys. Rev. Lett., 88, 101103

\bibitem[\protect\citeauthoryear{{Rheinhardt}, {Konenkov} \&
  {Geppert}}{{Rheinhardt} et~al.}{2004}]{2004A&A...420..631R}
{Rheinhardt} M.,  {Konenkov} D.,    {Geppert} U.,  2004, A\&A, 420, 631

\bibitem[\protect\citeauthoryear{{Shalybkov} \& {Urpin}}{{Shalybkov} \&
  {Urpin}}{1997}]{1997A&A...321..685S}
{Shalybkov} D.~A.,  {Urpin} V.~A.,  1997, A\&A, 321, 685

\bibitem[\protect\citeauthoryear{{Thompson} \& {Duncan}}{{Thompson} \&
  {Duncan}}{1995}]{1995MNRAS.275..255T}
{Thompson} C.,  {Duncan} R.~C.,  1995, MNRAS, 275, 255

\bibitem[\protect\citeauthoryear{{Thompson} \& {Duncan}}{{Thompson} \&
  {Duncan}}{1996}]{1996ApJ...473..322T}
{Thompson} C.,  {Duncan} R.~C.,  1996, ApJ, 473, 322

\bibitem[\protect\citeauthoryear{{Vainshtein}, {Chitre} \&
  {Olinto}}{{Vainshtein} et~al.}{2000}]{2000PhRvE..61.4422V}
{Vainshtein} S.~I.,  {Chitre} S.~M.,    {Olinto} A.~V.,  2000, Phys. Rev. E, 61, 4422

\bibitem[\protect\citeauthoryear{{Whitham}}{{Whitham}}{1974}]{1974G.B.Whitham}
{Whitham} G.~B.,  1974, {Linear and Nonlinear Waves}.
{Jhon Wiley, New York}

\end{thebibliography}

%
\end{document}